%% file: sn-article.tex
\documentclass[pdflatex,sn-mathphys-num]{sn-jnl}

\usepackage[T1]{fontenc}%
\usepackage[utf8]{inputenc}%
\usepackage{graphicx}%
\usepackage{multirow}%
\usepackage{amsmath,amssymb,amsfonts}%
\usepackage{amsthm}%
\usepackage[title]{appendix}%
\usepackage{xcolor}%
\usepackage{textcomp}%
\usepackage{manyfoot}%
\usepackage{booktabs}%
\usepackage{soul}

\theoremstyle{thmstyleone}%
\theoremstyle{thmstyletwo}%

\theoremstyle{thmstylethree}%

\begin{document}

\title[Article Title]{\textsc{JU\'A} — A Benchmark for Information Retrieval in Brazilian Legal Text Collections
}


\author[1,2]{\fnm{Jayr} \sur{Pereira}}\email{jayr.pereira@ufca.edu.br}

\author[3]{\fnm{Leandro Carísio} \sur{Fernandes}}\email{carisio@gmail.com}

\author[1]{\fnm{Erick} \sur{de Brito}}\email{erick.brito@aluno.ufca.edu.br}

\author[2,4]{\fnm{Roberto} \sur{Lotufo}}\email{roberto@neuralmind.ai}

\author[4]{\fnm{Luiz} \sur{Bonifacio}}\email{luiz.bonifacio@neuralmind.ai}


\affil*[1]{\orgdiv{Centro de Ciências e Tecnologia}, \orgname{Universidade Federal do Cariri (UFCA)}, \orgaddress{\city{Juazeiro do Norte}, \state{Ceará}, \country{Brazil}}}

\affil[2]{\orgdiv{Faculdade de Engenharia Elétrica e da Computação}, \orgname{Universidade Estadual de Campinas (UNICAMP)}, \orgaddress{\city{Campinas}, \state{São Paulo}, \country{Brazil}}}

\affil[3]{\orgname{Câmara dos Deputados}, \orgaddress{\city{Brasília}, \country{Brazil}}}

\affil[4]{\orgdiv{NeuralMind.ai}, \orgaddress{\city{Campinas}, \state{São Paulo}, \country{Brazil}}}


\abstract{Legal information retrieval in Portuguese remains difficult to evaluate systematically because available datasets differ widely in document type, query style, and relevance definition. We present \textsc{JU\'A}, a public benchmark for Brazilian legal retrieval designed to support more reproducible and comparable evaluation across heterogeneous legal collections. More broadly, \textsc{JU\'A} is intended not only as a benchmark, but as a continuous evaluation infrastructure for Brazilian legal IR, combining shared protocols, common ranking metrics, fixed splits, and a public leaderboard. The benchmark covers jurisprudence retrieval as well as broader legislative, regulatory, and question-driven legal search. We evaluate lexical, dense, and BM25-based reranking pipelines. Results show that the benchmark is sufficiently heterogeneous to distinguish retrieval paradigms and reveal substantial cross-dataset trade-offs: lexical search remains highly competitive on collections with strong lexical and institutional phrasing cues, while dense retrieval achieves statistically supported gains only on selected subsets, including question-driven matching. Overall, \textsc{JU\'A} provides a practical evaluation framework for studying legal retrieval across multiple Brazilian legal domains under a common benchmark design.}




\maketitle

\section{Introduction}\label{sec:introduction}

Legal information retrieval (legal IR) is a specialized area of information retrieval (IR) that focuses on retrieving relevant legal documents from a corpus given a user query \cite{lir_state_of_art_and_open_issues_SANSONE2022101967,concept_relevance_lir_vanOpijnen2017}. Although it shares the same general goal as traditional IR, legal IR operates in a domain with characteristics that make the task more challenging \cite{juristcu,normastcu,vitorio,br-tax-qa,nlp_legal_domain_ariai_2025}. Legal corpora can potentially cover all aspects of social life, resulting in large collections of documents that are typically long, institutionally heterogeneous, and strongly interconnected through citations and cross-references. They include statutes, case law, normative acts, and guidance materials produced for different audiences and purposes. Moreover, the legal system evolves continuously, with frequent updates to laws and regulations, introducing a temporal dimension that can affect retrieval \cite{lir_state_of_art_and_open_issues_SANSONE2022101967}.

Many legal IR applications must handle both legislative and judicial materials. Legislative texts tend to be formal and abstract \cite{lir_state_of_art_and_open_issues_SANSONE2022101967}, while judicial documents are often more narrative and combine facts, decisions, and legal reasoning in less rigid formats \cite{feng2024lcrsurvey}. In both cases, the language is technical, and legal relevance often differs from simple topical or semantic similarity \cite{concept_relevance_lir_vanOpijnen2017}. Documents with little lexical overlap may still be closely related when they share the same legal basis or precedent, whereas semantically similar expressions may have distinct legal consequences.

These challenges are important because retrieval is often the first step in legal processing pipelines, including e-discovery, compliance, contract analysis, and legal judgment prediction \cite{Qin_2024, Oard2010}. As in other multi-stage retrieval and retrieval-augmented pipelines, failures at this stage tend to propagate to downstream components, making retrieval quality critical for the reliability of legal AI systems \cite{karpukhin2020dpr,he2026llmlegaltasks,magesh2025hallucinationfree}.

This importance has not diminished with the rise of large language models (LLMs). Recent survey work on legal LLMs places legal IR among the core task families in LLM-based legal workflows, alongside document analysis, legal question answering, decision prediction, and legal text generation \cite{he2026llmlegaltasks}. This broader perspective reinforces that retrieval is not merely a supporting component, but part of the core infrastructure of legal AI. In practice, the growing use of retrieval-augmented generation makes retrieval quality even more consequential, since generative systems remain sensitive to the adequacy, coverage, and reliability of the legal evidence retrieved. Evaluations of leading AI legal research tools further suggest that retrieval and grounding failures remain a practical source of error, even in commercial systems that integrate retrieval with generation \cite{magesh2025hallucinationfree}.

Accordingly, progress in legal IR depends on the availability of datasets for systematic evaluation. Compared to general-purpose IR, legal datasets are scarcer, harder to obtain, and more expensive to annotate because they often require institutional access and domain expertise. Existing resources are also difficult to compare directly, since they vary in document type, query style, relevance definition, and evaluation setup. This problem is even more pronounced beyond English. In Portuguese, despite the recent emergence of valuable resources such as \textsc{JurisTCU}, \textsc{NormasTCU}, \textsc{Ulysses-RFCorpus}, and \textsc{BR-TaxQA-R}, there is still no unified public benchmark covering multiple legal retrieval regimes under a common evaluation framework \cite{juristcu,normastcu,vitorio,br-tax-qa}.

To address this gap, we propose \textsc{JU\'A}, which is, to the best of our knowledge, the first benchmark for legal IR across multiple types of Brazilian legal documents. The benchmark standardizes data organization and evaluation metrics, allowing direct comparison across retrieval paradigms. More importantly, we position \textsc{JU\'A} not only as a static benchmark release, but as a continuous evaluation infrastructure for Brazilian legal IR, centered on a public leaderboard and a shared submission protocol\footnote{\url{https://huggingface.co/spaces/ufca-llms/jua-leaderboard}}. \textsc{JU\'A} combines a new jurisprudence collection with adapted public legal IR resources under a shared evaluation design. The benchmark resources are publicly available on Hugging Face.

The main contributions of this research are twofold:
\begin{itemize}
\item We introduce a public, multi-domain benchmark for legal IR in Portuguese, encompassing multiple legal datasets, and a leaderboard that supports continuous, comparable evaluation.
\item We evaluate lexical, dense, and reranking strategies with independently fixed model checkpoints, illustrating how the benchmark exposes dataset-specific retrieval trade-offs under a shared protocol.
\end{itemize}

The remainder of this paper is organized as follows. Section~\ref{sec:related_work} reviews related work, Section~\ref{sec:methodology} describes the benchmark design and datasets, Section~\ref{sec:experiments} and Section~\ref{sec:results} present the experimental setup and results, and Section~\ref{sec:conclusions} concludes the paper.

\section{Related Work}\label{sec:related_work}

Legal IR differs from general-domain retrieval in at least three important ways. First, legal collections are highly structured and institutionally heterogeneous: jurisprudence, statutes, regulations, and administrative guidance differ in writing style, document length, granularity, and intended use. Second, relevance is often shaped by professional tasks such as precedent search, statutory interpretation, compliance checking, or question answering over official guidance, which means that different legal datasets may reward quite different retrieval behaviors. Third, legal language tends to combine formulaic expressions with domain-specific terminology and paraphrastic variation, making retrieval systems sensitive to both exact lexical overlap and more abstract semantic similarity. Recent survey work on legal case retrieval highlights this combination of structural complexity, specialized relevance notions, and strong dependence on task formulation \cite{feng2024lcrsurvey}.

For these reasons, legal IR has traditionally relied on lexical ranking methods such as BM25, which remain strong baselines because they are robust, interpretable, and highly effective when query terms overlap with legally salient document expressions \cite{robertson2009bm25}. More recent dense retrieval methods address lexical mismatch by mapping queries and documents into a shared embedding space, allowing semantic matching beyond exact term overlap \cite{karpukhin2020dpr}. This transition from sparse to dense retrieval is particularly relevant in legal collections, where synonymous formulations, paraphrases, and document-specific drafting conventions can make pure lexical matching insufficient.

Furthermore, in Brazilian Portuguese, the main bottleneck is not only modeling, but also evaluation. Recent resources such as \textsc{JurisTCU} \cite{juristcu}, \textsc{NormasTCU} \cite{normastcu}, the Ulysses relevance feedback corpus \cite{vitorio}, and \textsc{BR-TaxQA-R} \cite{br-tax-qa} have broadened the empirical basis for legal NLP and legal retrieval in Brazilian Portuguese. These resources are individually valuable, although they target different institutions, query types, and annotation procedures. As a result, they improve domain coverage without providing a unified benchmark for direct comparison across retrieval paradigms. More broadly, Portuguese IR research has also benefited from recent public datasets such as Quati \cite{quati}, but legal-domain retrieval remains comparatively under-benchmarked and more fragmented.

This motivates a benchmark perspective closer to heterogeneous IR evaluation frameworks such as BEIR \cite{thakur2021beir}, which showed the value of testing retrieval methods across multiple datasets under a common protocol, and MTEB \cite{muennighoff2022mteb}, which reinforced the importance of shared evaluation infrastructure for embedding-based models. A similar lesson also emerges from legal-domain shared tasks such as AILA \cite{bhattacharya2019aila} and COLIEE \cite{rabelo2022coliee}, where task design and common evaluation procedures are central to making progress comparable across systems. In both the general and legal IR settings, the broader lesson is that evaluation quality depends not only on the number of datasets available but also on the extent to which they are made comparable under common metrics, fixed splits, and transparent reporting. \textsc{JU\'A} follows this general philosophy, but specializes it for Brazilian legal retrieval: rather than introducing a single new dataset, it combines multiple legal retrieval settings into a shared benchmark with common reporting and a public leaderboard.

\section{Benchmark Design and Datasets}\label{sec:methodology}

\textsc{JU\'A} is designed to evaluate information retrieval systems in the Brazilian legal domain under a unified and reproducible protocol. The \textsc{JU\'A} benchmark combines complementary datasets spanning legal and regulatory domains, covering jurisprudence, normative text, and question-driven retrieval settings. \textsc{JU\'A-Juris} and \textsc{JurisTCU} evaluate search over jurisprudence collections, while \textsc{NormasTCU}, \textsc{Ulysses-RFCorpus}, and \textsc{BR-TaxQA} cover broader legislative, regulatory, or legal question--answer retrieval scenarios. 

\textsc{JU\'A-Juris} is newly introduced in this work; the remaining four datasets are adapted or repackaged from existing public legal IR resources under the benchmark's common evaluation protocol. Table~\ref{tab:jua_dataset_summary} summarizes the role of each dataset in the benchmark. The benchmark can be naturally extended with new legal collections in future releases and is paired with a public leaderboard, which continuously ranks submissions under fixed evaluation settings and supports transparent comparison over time. The name \textsc{JU\'A} references the \textit{ju\'a} fruit, which is common in Northeastern Brazil, connecting the benchmark identity to its regional and linguistic context.

This mixed-regime design is intentional. Rather than modeling Brazilian legal retrieval as a single homogeneous task, \textsc{JU\'A} provides a common evaluation framework for a set of complementary legal retrieval settings. Jurisprudence search, normative retrieval, legislative search, and question-driven legal retrieval differ in document structure, query formulation, and relevance definition, and these differences make a unified benchmark informative. In this design, comparability comes from a shared evaluation protocol rather than from an assumption that all subsets instantiate the same retrieval problem. By preserving per-dataset reporting alongside aggregate summaries, \textsc{JU\'A} supports both direct comparison and regime-sensitive interpretation.

\input{tables/dataset_summary}

\subsection{\textsc{JU\'A-Juris}: TCU Curated Jurisprudence}

\textsc{JU\'A-Juris} is derived from the curated jurisprudence collection maintained by the Brazilian Federal Court of Accounts (TCU). In this collection, each record pairs an \textit{enunciado}, an abstractive summary of a legal holding, with an \textit{excerto}, the passage in the decision that supports that summary. The collection is a curated subset of TCU precedents considered legally salient, including novel holdings, decisions with substantial collegial discussion, and reiterated institutional understandings \cite{tcu_jurisprudencia_portal,tcu_dados_abertos_jurisprudencia}. Thus, the task is specifically the retrieval of a supporting legal excerpt from an abstractive summary, rather than a direct simulation of a lawyer's or citizen's natural search query.

We obtained the collection through the TCU open-data export\footnote{\url{https://sites.tcu.gov.br/dados-abertos/}} as a pipe-separated tabular file. We retained the record key, the \texttt{ENUNCIADO} and \texttt{EXCERTO} fields, and the decision number and year used to create a human-readable title. Preprocessing removed HTML markup, normalized \textit{s\'umula}-style prefixes in the summaries, and excluded rows with missing summaries, excerpts, decision identifiers, or placeholder excerpt text. This procedure yielded 17{,}147 eligible query--excerpt pairs. The \textit{enunciado} is the query, the corresponding \textit{excerto} is the positive passage, and the record key identifies both sides of the paired relevance judgment.

We use a 90/10 train--test partition: 15{,}433 pairs for training and 1{,}714 for testing. Test-set selection uses the \emph{BM25 positive-document rank}, a BM25-specific sampling signal. For each source query, we record the rank at which BM25 retrieves its paired item in the original candidate list. Rank 1 denotes an easier case for BM25; higher ranks indicate that BM25 needs to inspect more candidates; and \texttt{not\_found} indicates that the paired item is absent from its top 1,000 candidates. This signal is not an intrinsic measure of query difficulty and does not predict the difficulty that another retriever will face.

The test set was sampled across logarithmic rank bins to retain cases with different degrees of lexical accessibility: \texttt{bin\_1}: 67; \texttt{bin\_2}: 16; \texttt{bin\_4}: 89; \texttt{bin\_8}: 145; \texttt{bin\_16}: 119; \texttt{bin\_32}: 326; \texttt{bin\_64}: 308; \texttt{bin\_128}: 203; \texttt{bin\_256}: 131; \texttt{bin\_512}: 94; \texttt{bin\_1024}: 73; and \texttt{not\_found}: 143. The bins make the test set useful for analyzing robustness beyond cases that are immediately retrieved by a lexical baseline, but they do not establish a universal ordering of difficulty across retrieval models.

The qrels encode one binary paired positive per query (\texttt{query-id == corpus-id}). They are therefore not exhaustive judgments of every decision that could be legally relevant: another decision may express the same principle without being annotated as relevant. We treat this as a limitation of the current release and identify pooled assessment of retrieved candidates by legal experts as future work for multi-positive relevance judgments. The dataset is publicly available on Hugging Face\footnote{\url{https://huggingface.co/datasets/ufca-llms/jua}}.

\subsection{\textsc{JurisTCU}: TCU Curated Jurisprudence}

\textsc{JurisTCU} is a Brazilian Portuguese legal IR resource built from the curated jurisprudence collection of the Brazilian Federal Court of Accounts (TCU) \cite{juristcu}, from which we derive the benchmark subset used here.
In its original form, the dataset contains 16{,}045 jurisprudence documents organized into more than 20 metadata and textual fields. Our adaptation uses the \texttt{EXCERTO} field as the retrieval corpus.

The dataset also has 150 queries with relevance judgments, which include real keyword queries extracted from user searches, synthetic keyword queries generated from the summaries of the most accessed documents (\textit{enunciado}), and synthetic question-based queries generated from these summaries. Relevance judgments were initially produced using an LLM and subsequently verified manually.

In our \textsc{JU\'A} adaptation, we kept only the \texttt{EXCERTO} field, which contains, on average, approximately 4,400 characters (660 words). Because some queries were generated from the \texttt{ENUNCIADO} field, this choice intentionally makes retrieval more challenging for both lexical and semantic models, since the absence of a concise summary makes it harder to capture the document’s core terms and semantics.

Within \textsc{JU\'A}, this dataset is the second jurisprudence-focused retrieval setting and complements \textsc{JU\'A-Juris} with keyword, synthetic, and question-style queries plus expert-verified relevance judgments. Together with the more user-oriented queries in \textsc{Ulysses-RFCorpus} and \textsc{BR-TaxQA}, it provides a complementary counterpoint to the abstractive-summary formulation of \textsc{JU\'A-Juris}. The adapted dataset is available on Hugging Face\footnote{\url{https://huggingface.co/datasets/ufca-llms/juris-tcu}}.

\subsection{\textsc{NormasTCU}: TCU Normative Texts}

\textsc{NormasTCU} is a Brazilian Portuguese legal IR test collection composed of normative documents from the Brazilian Federal Court of Accounts (TCU) \cite{normastcu}. These normative acts may have internal effects (e.g., rules governing internal procedures) or external effects (e.g., rules regulating how the court interacts with other public institutions) and differ from jurisprudential documents in both purpose and structure. Jurisprudential documents typically describe specific cases and present the legal reasoning that leads to a decision, whereas normative acts establish general rules that must be applied to specific cases. Moreover, they exhibit a distinct textual structure: jurisprudential documents are usually organized as narrative legal arguments, while normative acts are structured hierarchically into articles, paragraphs, and other legal provisions that define prescriptive rules.

In its original form, \textsc{NormasTCU} contains 14{,}469 legal documents. Each document corresponds to a normative act and is structured into multiple metadata and textual fields. Among these, the most relevant are \texttt{TEXTONORMA}, which contains the full text of the normative act, and \texttt{ASSUNTO}, a summary describing its content. The \texttt{ASSUNTO} field can help users and retrieval systems identify documents relevant to a given query. However, this field is missing in more than 7{,}000 documents.

The collection includes 46 queries, some expressed as short keyword queries (less informative and closer to real user searches) and others written in a more semantically rich form. Relevance judgments were produced by four domain experts using a three-level graded scale. In total, the qrels contain assessments for 812 unique judged query--document pairs.
For the \textsc{JU\'A} benchmark, we represent each document by concatenating the \texttt{ASSUNTO} and \texttt{TEXTONORMA} fields, combining the norm summary with its full text. The dataset is publicly available on Hugging Face\footnote{\url{https://huggingface.co/datasets/ufca-llms/normas-tcu}}.

\subsection{Ulysses-RFCorpus}

\textsc{Ulysses-RFCorpus} is a Brazilian Portuguese corpus for legislative legal IR with explicit relevance feedback collected in a real production scenario at the Brazilian Chamber of Deputies \cite{vitorio}. The corpus was designed to capture user feedback from the institution's own retrieval workflow, rather than synthetic labels.

The Brazilian Chamber of Deputies includes a specialized department, Legislative Consulting, which supports parliamentarians during the law-making process. In this workflow, parliamentarians can request this department to draft new bills. The queries in \textsc{Ulysses-RFCorpus} correspond to these requests, anonymized, and reflect real information needs. The document collection is composed of legislative proposals (bills). The relevance judgments were provided by 54 legislative consultants using a three-level scale over a pool of retrieved documents, and they could also manually indicate additional relevant bills.

According to the authors, the resource is especially relevant because publicly available legal corpora with expert or user-provided relevance feedback are still scarce, particularly in the legislative domain \cite{vitorio}. The collection includes 692 queries with relevance judgments, reflecting diverse legislative information needs and real user feedback patterns captured during operational retrieval sessions. In this sense, \textsc{Ulysses-RFCorpus} complements predominantly judicial benchmarks by providing supervision aligned with legislative drafting and policy-support tasks.

Within \textsc{JU\'A}, this subset brings relevance judgments grounded in the real operational needs of legislative consultants. This improves ecological validity and helps evaluate whether models generalize beyond lexical matching to the kind of judgments observed in practical legal-government use. The adapted dataset is publicly available on Hugging Face\footnote{\url{https://huggingface.co/datasets/ufca-llms/Ulysses-RFCorpus}}.

\subsection{\textsc{BR-TaxQA-R}: Federal Revenue Service Q\&A}

\textsc{BR-TaxQA-R} \cite{br-tax-qa} is a collection derived from materials of the Brazilian Federal Revenue Service (\textit{Receita Federal}) on personal income tax (IRPF). It contains 715 questions with their corresponding answers and introduces user-oriented, FAQ-style query formulations in a legal-tax domain. We refer to its adapted benchmark subset as \textsc{BR-TaxQA} throughout the remainder of the paper.

Some questions are explicitly linked to other related questions. In our relevance design, the immediate answer to the queried question is treated as the primary positive with score $=2$, while linked answers are treated as secondary positives with score $=1$. This setup enables graded relevance evaluation and better reflects practical retrieval behavior in interlinked FAQ structures. This dataset has been used for question-answering tasks in other works, such as the one described in \cite{brito2025avaliando}. The adapted dataset is publicly available on Hugging Face\footnote{\url{https://huggingface.co/datasets/ufca-llms/BR-TaxQA}}.

\section{Experiments} \label{sec:experiments}

This section describes our experimental setup for evaluating retrieval methods on \textsc{JU\'A} benchmark. The goal of these experiments is not only to rank systems according to their retrieval effectiveness under the benchmark metrics, but also to validate that the benchmark distinguishes retrieval behaviors across legal subdomains. We therefore present the model families considered in the benchmark and the ranking metrics used to compare systems, all under a unified, reproducible protocol. Because the benchmark combines two jurisprudence-search datasets (\textsc{JU\'A-Juris} and \textsc{JurisTCU}) with three broader legislative, regulatory, or question-driven retrieval datasets (\textsc{NormasTCU}, \textsc{Ulysses-RFCorpus}, and \textsc{BR-TaxQA}), the experiments assess dataset-specific behavior and cross-setting robustness.

\subsection{Models}

We evaluate first-stage retrieval models and reranking variants.
BM25 is used as the main lexical baseline for comparison with dense and rerank models.
Model selection was guided by the top-performing entries on the public MTEB leaderboard, which has become a standard reference for broad embedding evaluation \cite{muennighoff2022mteb}. The primary comparison includes only models with externally fixed, publicly released checkpoints.

\textbf{First-stage retrieval models}:
\begin{itemize}
    \item \textbf{BM25}: sparse lexical baseline.
    \item \textbf{Qwen3-Embedding dense retrievers}: Qwen3-Embedding-0.6B, Qwen3-Embedding-4B, and Qwen3-Embedding-8B \cite{qwen3embedding}.
    \item \textbf{Harrier OSS v1 dense retrievers}: Harrier OSS v1-270M, Harrier OSS v1-0.6B, and Harrier OSS v1-27B \cite{microsoft2026harrier}. Harrier OSS v1-27B was a state-of-the-art MTEB model when the comparison set was selected.
    \item \textbf{Closed commercial dense baseline}: OpenAI 3-small.
\end{itemize}

All dense retrievers were evaluated with the same prompt-free encoding procedure: we passed the raw query text and did not apply model-specific query instructions. This establishes a uniform evaluation condition, although it does not represent the maximum instructed performance of models whose documentation recommends task-specific prompts.

\textbf{Reranking models}: we also evaluate rerank configurations for Qwen3-Embedding-8B, Qwen3-Embedding-4B, OpenAI 3-small, Harrier OSS v1-27B, and Harrier OSS v1-0.6B, reported separately in the same leaderboard table. In the reranking setup, each model reorders the top-1000 candidate documents returned by BM25 for each query.

\subsection{Metrics}

Leaderboard reporting follows four ranking metrics at a cutoff of 10:
\begin{itemize}
    \item \textbf{NDCG@10}: position-aware ranking quality that discounts lower-ranked hits and is especially useful when relevance is graded rather than purely binary \cite{jarvelin2002cumulated}.
    \item \textbf{MRR@10}: emphasizes the rank of the first relevant hit, making it particularly informative in settings where queries have only one or very few relevant documents \cite{voorhees2000building}.
    \item \textbf{MAP@10}: summarizes ranking quality by averaging precision at the ranks where relevant documents are retrieved, giving a broader view of top-ranked effectiveness than first-hit metrics alone \cite{manning2008introduction}.
    \item \textbf{Recall@10}: measures how much of the relevant material is recovered within the top 10 results, highlighting top-10 coverage rather than only rank ordering \cite{manning2008introduction}.
\end{itemize}

Table~\ref{tab:leaderboard_models} reports dataset-wise NDCG@10 and MRR@10 scores, together with an overall score computed as the mean across available datasets for each model. We use MAP@10 and Recall@10 for complementary visual analysis in the Results section, as these metrics help distinguish early-hit behavior from broader top-10 ranking quality and coverage. The public leaderboard also tracks P@10, but we do not emphasize it in our results because MAP@10 and Recall@10 provide a clearer complementary view of retrieval behavior across heterogeneous legal retrieval settings in the benchmark configuration studied here.

\section{Results} \label{sec:results}

\subsection{Overall Benchmark Behavior}

Table~\ref{tab:leaderboard_models} shows no dominant retrieval paradigm across \textsc{JU\'A}: lexical, dense, and reranking systems exchange positions depending on the dataset, and none is uniformly superior. This lack of a single expressive winner is itself informative, because it suggests that the benchmark is not saturated by one retrieval strategy and that retrieval behavior remains sensitive to differences in legal text type, query formulation, and relevance structure.

At the aggregate level, the benchmark also shows that strong overall performance does not imply uniform superiority across legal retrieval settings. This is precisely the kind of behavior a heterogeneous benchmark should reveal: models that look strong on the overall average may still depend on particular task regimes, document types, or relevance structures. This pattern is consistent with findings from heterogeneous retrieval benchmarks such as BEIR, where BM25 remains a robust baseline rather than a trivial one \cite{thakur2021beir}. A concrete example is the argument-retrieval subset Touch\'e 2020, a task that has proven difficult even for strong neural retrievers and where subsequent analysis found them still less effective than BM25 even after dataset denoising \cite{thakur2024touche}. In this sense, \textsc{JU\'A} is useful not only for ranking systems, but also for exposing where retrieval strategies remain sensitive to the structure of the underlying legal collection.

\input{tables/leaderboard}

\subsection{Dataset-Specific Analysis}

The dataset-level results are easier to interpret when read alongside the design of the benchmark. In practice, they point to two broad groups of retrieval tasks within \textsc{JU\'A}: jurisprudence search on one side, and broader legislative, regulatory, or question-driven retrieval on the other. Within the jurisprudence group, however, the results are not uniform. In Table~\ref{tab:leaderboard_models}, Harrier OSS v1-27B is the strongest independently fixed dense retriever on \textsc{JU\'A-Juris} (row 4), while the lexical baseline remains meaningful because the \textit{enunciados} are concise summaries of the legal holding expressed in the corresponding decisions. This formulation increases lexical consistency between queries and relevant jurisprudential passages and naturally benefits exact-match retrieval. On the other hand, Table~\ref{tab:leaderboard_models} shows that BM25 is the strongest first-stage retrieval system on \textsc{JurisTCU} (row 1), indicating that legal query wording and relevant excerpts still share enough salient lexical structure for sparse retrieval to remain highly competitive. Taken together, these results suggest that ``jurisprudence retrieval'' should not be treated as a single regime within \textsc{JU\'A}: even its two jurisprudence-focused subsets reward different retrieval behaviors.

The remaining datasets are closer to broader legislative, regulatory, or question-driven legal search. On \textsc{Ulysses-RFCorpus}, BM25 is again highly competitive (row 1 in Table~\ref{tab:leaderboard_models}), which likely reflects the importance of exact term overlap, recurring institutional phrasing, and domain-specific lexical patterns in legislative retrieval. \textsc{NormasTCU} occupies a more difficult middle ground, where no approach is clearly dominant in the available comparisons and where both sparse and dense methods must cope with long normative texts, article-level structure, and highly standardized regulatory language. Figure~\ref{fig:significance_matrices} shows no statistically significant difference from BM25 for any model with complete per-query coverage on this subset. The current evidence therefore does not establish a consistent advantage for either retrieval family. By contrast, \textsc{BR-TaxQA} provides the clearest evidence of a dense-retrieval advantage: OpenAI 3-small, Harrier OSS v1-27B, and both Harrier OSS v1 rerankers significantly outperform BM25 on both metrics (Figure~\ref{fig:significance_matrices}), with Harrier OSS v1-27B and its reranking variant providing the leading dense point estimates (row 4 and row 11 in Table~\ref{tab:leaderboard_models}). This contrast is useful for the benchmark: it shows that the non-jurisprudential portion of \textsc{JU\'A} is also heterogeneous, spanning regulatory retrieval tasks closer to document search and FAQ-style retrieval tasks closer to semantic answer matching.

\subsection{Model Scale and Family Comparison}

The first-stage Harrier OSS v1 family permits a descriptive comparison across three scales under a common prompt-free protocol. We also include Qwen3-Embedding-0.6B, Qwen3-Embedding-4B, and Qwen3-Embedding-8B as general-purpose reference points. Because the two model families differ in architecture and training data, this is not a controlled cross-family scaling experiment; lines in the figure connect only models from the same family.

\begin{figure*}[t]
\centering
\includegraphics[width=0.98\textwidth]{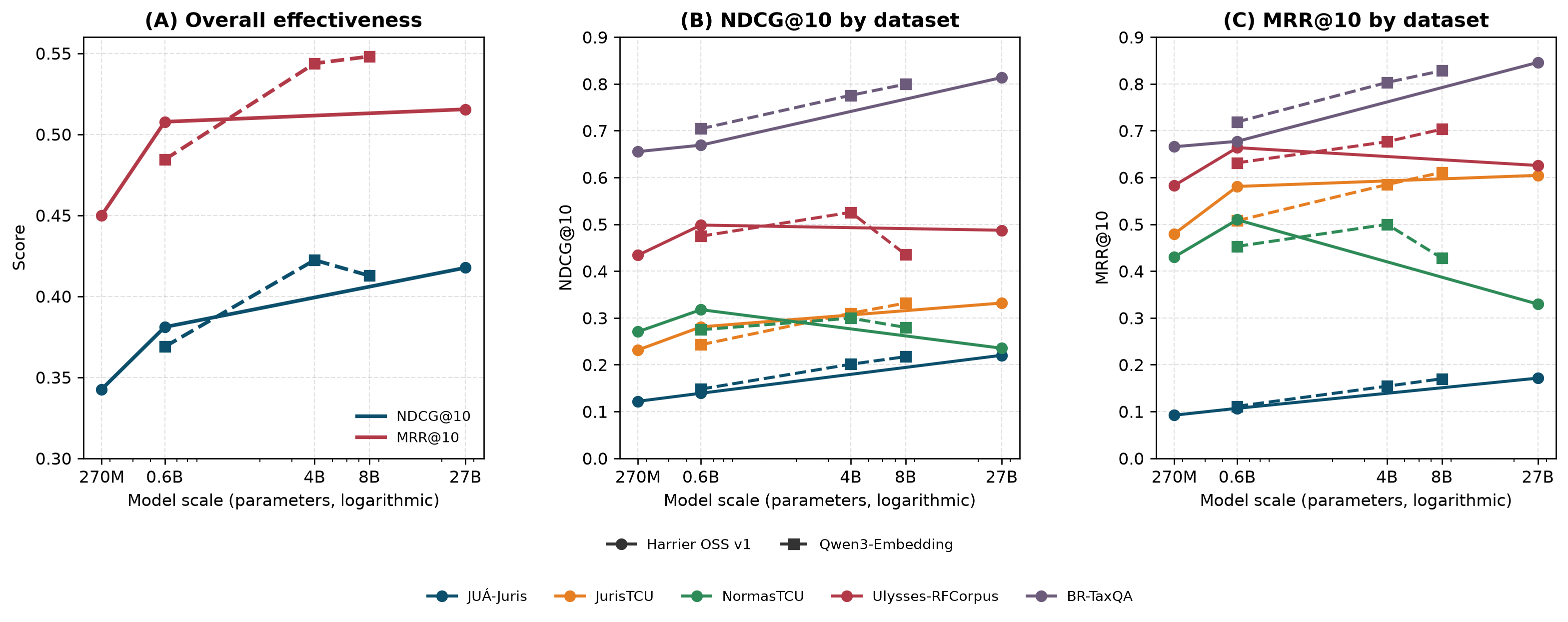}
\caption{First-stage retrieval effectiveness by model scale and model family. Solid lines represent Harrier OSS v1 and dashed lines represent Qwen3-Embedding; lines connect models within a family only. Harrier OSS v1 overall NDCG@10 increases from 0.343 (270M) to 0.381 (0.6B) and 0.418 (27B), while its overall MRR@10 increases from 0.450 to 0.508 and 0.515. The Qwen3-Embedding reference points show that parameter count alone does not determine performance across families.}
\label{fig:harrier_scaling}
\end{figure*}

Figure~\ref{fig:harrier_scaling} summarizes the within-family Harrier OSS v1 scaling trend and situates it relative to Qwen3-Embedding. For Harrier OSS v1, aggregate NDCG@10 improves monotonically with scale. Its MRR@10 also rises sharply from 270M to 0.6B, with a smaller additional increase at 27B. Qwen3-Embedding shows a similar initial gain: overall NDCG@10 and MRR@10 increase from 0.369 and 0.484 at 0.6B to 0.422 and 0.544 at 4B. Moving from 4B to 8B slightly lowers NDCG@10 (0.413 versus 0.422) while improving MRR@10 (0.548 versus 0.544). Thus, parameter count alone does not explain cross-model performance.

The dataset-level trajectories further qualify the aggregate pattern. Within Harrier OSS v1, the 27B model is strongest on \textsc{JU\'A-Juris}, \textsc{JurisTCU}, and \textsc{BR-TaxQA}, whereas the 0.6B model performs best on \textsc{NormasTCU} and \textsc{Ulysses-RFCorpus}. The Qwen3-Embedding trajectory is likewise non-uniform across datasets. Scaling can therefore improve aggregate effectiveness within a family without producing the best result for every collection or a uniform ordering across model families.

\subsection{Statistical Comparison}
\label{sec:statistical_comparison}

\begin{figure*}[t]
\centering
\includegraphics[width=0.98\textwidth]{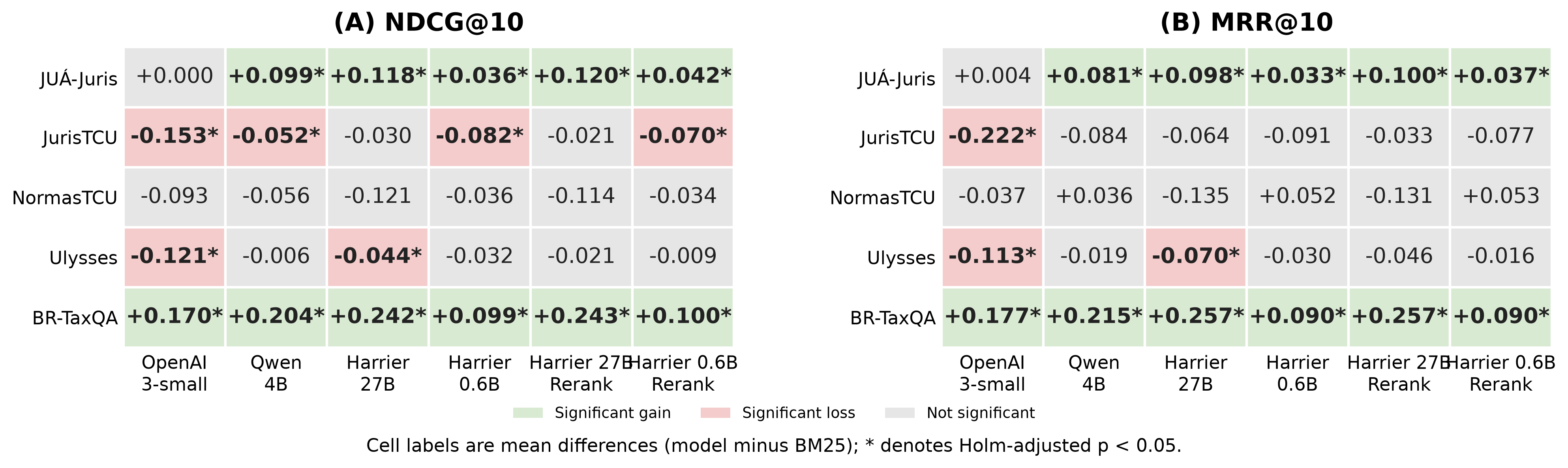}
\caption{Paired comparisons against BM25 for models with complete per-query runs on all five datasets: (A) NDCG@10 and (B) MRR@10. Each cell reports the mean per-query difference between the model and BM25. Green and red cells denote a significant gain and loss, respectively, after Holm correction; gray cells are not significant.}
\label{fig:significance_matrices}
\end{figure*}

To determine which point-estimate differences are robust at the query level, we compare BM25 with dense retrievers and rerankers whenever complete per-query runs are available and reproduce the corresponding Table~\ref{tab:leaderboard_models} values. We compute paired-bootstrap 95\% confidence intervals (5,000 resamples), two-sided paired randomization tests (20,000 sign permutations), and Holm correction across 64 available metric-level comparisons. Figure~\ref{fig:significance_matrices} reports the models with complete per-query coverage across all five datasets, including Qwen3-Embedding-4B; the supplementary material reports all available comparisons, including non-significant tests and partial-coverage Qwen3-Embedding-8B and Qwen reranking comparisons.

The inferential results provide a direct answer to where dense retrieval beats lexical search: statistically supported gains occur in two subsets, not across the benchmark as a whole. On \textsc{JU\'A-Juris}, Qwen3-Embedding-4B, both Harrier OSS v1 first-stage models, and both Harrier rerankers significantly outperform BM25 on both NDCG@10 and MRR@10, with mean gains ranging from $+0.033$ to $+0.120$. On \textsc{BR-TaxQA}, OpenAI 3-small, Qwen3-Embedding-4B, both Harrier OSS v1 first-stage models, and both Harrier rerankers significantly outperform BM25 on both metrics; Harrier OSS v1-27B and its reranking variant yield the largest observed effects (up to $+0.243$ NDCG@10 and $+0.257$ MRR@10). These results are compatible with the semantic matching demands of question--answer retrieval in \textsc{BR-TaxQA} and show that dense representations can also improve summary--excerpt retrieval in \textsc{JU\'A-Juris}; however, they do not isolate a single causal explanation for either gain.

The remaining comparisons delimit this conclusion. BM25 significantly outperforms OpenAI 3-small on \textsc{JurisTCU} and \textsc{Ulysses-RFCorpus}, Harrier OSS v1-27B on \textsc{Ulysses-RFCorpus}, and Qwen3-Embedding-4B and Harrier OSS v1-0.6B on NDCG@10 for \textsc{JurisTCU}; no \textsc{NormasTCU} comparison in Figure~\ref{fig:significance_matrices} remains significant after Holm correction, consistent with its small test set and the uncertainty visible in the confidence intervals. Harrier OSS v1-0.6B Rerank also has lower NDCG@10 on \textsc{JurisTCU}, despite its gains on \textsc{JU\'A-Juris} and \textsc{BR-TaxQA}. The evidence therefore supports conditional, dataset-specific claims about the benefit of dense retrieval and reranking rather than a universal advantage in Brazilian legal IR.

The comparison with OpenAI 3-small, a closed commercial embedding model, is particularly informative: it significantly outperforms BM25 on \textsc{BR-TaxQA}, but BM25 significantly outperforms it on \textsc{JurisTCU} and \textsc{Ulysses-RFCorpus}. Thus, closed-model access alone does not yield a uniform advantage across Brazilian legal IR tasks. Harrier OSS v1-27B, a state-of-the-art MTEB model at selection time \cite{microsoft2026harrier}, shows the same benchmark-dependent pattern: it yields significant gains on \textsc{JU\'A-Juris} and \textsc{BR-TaxQA}, while BM25 has a significant advantage on \textsc{Ulysses-RFCorpus}. These comparisons illustrate why a multi-dataset benchmark is needed even when evaluating strong general-purpose embedding models.

\subsection{Effect of Reranking}

Among first-stage dense retrievers, Qwen3-Embedding-4B obtains the strongest overall NDCG@10 (0.422; row 3 in Table~\ref{tab:leaderboard_models}), whereas Qwen3-Embedding-8B obtains the strongest overall MRR@10 (0.548; row 2). This distinction reinforces that aggregate effectiveness depends on the metric: the 4B model provides stronger position-aware ranking quality, while the 8B model offers slightly better early-hit behavior. BM25 (row 1) nevertheless remains a strong baseline and should not be treated as a weak point of comparison.

Reranking the top-1000 BM25 candidates produces higher point estimates for some configurations, although not uniformly. Qwen3-Embedding-4B Rerank and Harrier OSS v1-27B Rerank share the highest overall NDCG@10 point estimate (0.426; rows 10 and 11), while Qwen3-Embedding-8B Rerank obtains the strongest overall MRR@10 (0.558; row 9). Harrier OSS v1-0.6B Rerank produces the strongest reranking point estimates on \textsc{NormasTCU} (0.322 NDCG@10 and 0.517 MRR@10; row 12). On \textsc{Ulysses-RFCorpus}, Qwen3-Embedding-4B Rerank obtains the strongest reranking NDCG@10 (0.537; row 10), while Qwen3-Embedding-8B Rerank obtains the strongest MRR@10 (0.716; row 9). A direct paired comparison between Qwen3-Embedding-4B Rerank and its first-stage counterpart finds a statistically significant gain only for NDCG@10 on \textsc{Ulysses-RFCorpus} ($+0.011$, 95\% CI $[+0.007, +0.016]$, Holm-adjusted $p=0.0015$); the remaining metric-level differences are not significant after correction. Both Harrier OSS v1 rerankers now have complete per-query coverage across the benchmark and significantly improve on BM25 for \textsc{JU\'A-Juris} and \textsc{BR-TaxQA}; Harrier OSS v1-0.6B Rerank also has lower NDCG@10 on \textsc{JurisTCU} (Figure~\ref{fig:significance_matrices}). Complete per-query rankings remain unavailable for the Qwen3-Embedding-8B and OpenAI reranking configurations on one or more datasets. Their aggregate scores can therefore be reported in Table~\ref{tab:leaderboard_models}, but we cannot recompute paired confidence intervals or statistical tests against BM25 across the full benchmark; these values are not used as inferential evidence of a reranking improvement. Qwen3-Embedding-4B Rerank is instead evaluated directly against its first-stage counterpart above and is omitted from Figure~\ref{fig:significance_matrices}, which focuses on comparisons against BM25.

Table~\ref{tab:reranker_comparison} isolates the effect of reranking by comparing each reranker with the first-stage version of the same model. Qwen3-Embedding-4B gains significantly only on NDCG@10 for \textsc{Ulysses-RFCorpus}. Harrier OSS v1-27B gains significantly on both metrics for the same subset, while its remaining changes are not significant after correction. Harrier OSS v1-0.6B gains on NDCG@10 for \textsc{JU\'A-Juris}, on both metrics for \textsc{JurisTCU}, and on both metrics for \textsc{Ulysses-RFCorpus}. Thus, reranking improvements are model- and collection-dependent rather than a general property of the pipeline.

\input{tables/reranker_comparison}

When BM25 already produces a strong candidate ordering because of controlled vocabulary, recurrent institutional phrasing, or highly diagnostic legal terminology, reranking has limited headroom and may even perturb an already well-ordered list. Reranking is therefore best viewed as a complementary pipeline component whose benefits depend on the candidate set and the dataset, rather than as a guaranteed improvement in every setting.

Taken together, these results show that \textsc{JU\'A} exposes substantial cross-dataset trade-offs and is useful for distinguishing retrieval behavior beyond a single legal collection.

\subsection{Qualitative Error Analysis}

To better understand the quantitative patterns above, we also inspected query-level runs for \textsc{JU\'A-Juris}, \textsc{JurisTCU}, and \textsc{Ulysses-RFCorpus}. The examples below are illustrative rather than estimates of prevalence or effect magnitude: Figure~\ref{fig:significance_matrices} and the supplementary material provide the corresponding query-level effect sizes, confidence intervals, and corrected significance tests where complete paired runs are available. Even this small qualitative sample helps clarify that the benchmark does not consist solely of one type of difficulty. Some queries are easy for both lexical and dense retrieval, others are difficult for both, and a third group sharply separates sparse and dense behavior.

Easy cases are short, topically focused, and lexically stable. In \textsc{JurisTCU}, queries such as ``\textit{t\'ecnica e pre\c{c}o}'', ``\textit{auditoria interna}'', and ``\textit{padroniza\c{c}\~ao marca}'' are retrieved at rank 1 by both BM25 and the dense retriever. In \textsc{Ulysses-RFCorpus}, both methods also solve direct legislative queries such as ``\textit{defesa da concorr\^encia}'', ``\textit{exporta\c{c}\~ao de commodities}'', and ``\textit{proibir refrigerante escolas}'' very early. These examples suggest that when the information need is expressed in concise, institutionally familiar wording, both paradigms can exploit the signal effectively.

Hard cases, by contrast, often combine abstraction, long-form formulation, or weaker lexical anchoring. In \textsc{JU\'A-Juris}, even apparently well-formed statements such as ``\textit{\'E irregular a aplica\c{c}\~ao de recursos fora da vig\^encia do conv\^enio}'' remain outside the top 10 for both BM25 and dense retrieval in our inspected runs. In \textsc{Ulysses-RFCorpus}, longer requests submitted to the Legislative Consulting department, such as a proposal to make the right to life a constitutional, immutable clause or a request to modernize title-credit legislation, also remain difficult for both methods. These examples suggest that lexical clarity alone is insufficient when the retrieval target depends on more diffuse legal framing or broader argumentative context.

The contrast between BM25 and dense retrieval is especially informative. BM25 tends to win when the query contains highly diagnostic legal surface forms, statutory references, or institutional terminology. In \textsc{JU\'A-Juris}, for example, the query beginning ``\textit{O \'org\~ao ou a entidade concedente somente deve firmar conv\^enios...}'' is retrieved at rank 1 by BM25 but only at rank 346 by the dense model. In \textsc{JurisTCU}, brief legal cues such as ``\textit{decreto-lei 4.657/1942}'' and ``\textit{novo e improrrog\'avel prazo}'' show a similar pattern. In \textsc{Ulysses-RFCorpus}, BM25 also benefits from explicit statutory or domain-specific wording, as in ``\textit{alterar a Lei 12.249...}'' and ``\textit{uso de bioinsumos no Brasil}'', where exact lexical anchoring appears to dominate. At the aggregate level, this pattern is supported for OpenAI 3-small: Figure~\ref{fig:significance_matrices} shows significant BM25 advantages on both metrics in \textsc{JurisTCU} and \textsc{Ulysses-RFCorpus}, with negative dense-minus-BM25 effects from $-0.113$ to $-0.222$.

Dense retrieval can win under a different regime. It performs especially well when the query is phrased more naturally, paraphrased, or exhibits weaker exact lexical overlap with the relevant document. In \textsc{JU\'A-Juris}, the query ``\textit{N\~ao constitui requisito pr\'evio \`a contrata\c{c}\~ao a disponibilidade de recursos financeiros...}'' is ranked first by the dense model, while BM25 fails to retrieve the gold document in the inspected candidate list. This type of case is consistent with the significant dense-retrieval gains on \textsc{JU\'A-Juris} in Figure~\ref{fig:significance_matrices}, whose positive dense-minus-BM25 effects range from $+0.037$ to $+0.118$. In \textsc{JurisTCU}, short but semantically abstract queries such as ``\textit{medida cautelar}'' and ``\textit{controle ades\~ao}'' favor the dense model, as does the question-like query ``\textit{\'E obrigat\'oria pesquisa de mercado na elabora\c{c}\~ao do or\c{c}amento-base da licita\c{c}\~ao?}''. In \textsc{Ulysses-RFCorpus}, dense retrieval is particularly strong on natural-language legislative requests such as regulating Airbnb-like hosting platforms, renegotiating state debt with the Union, or regulating tele-physical-education services. These individual cases do not imply aggregate dense-retrieval gains on those datasets: the paired OpenAI and Harrier OSS v1-27B comparisons favor BM25 in both metrics. Conversely, the statistically supported gains of OpenAI 3-small, Harrier OSS v1-27B, and Harrier OSS v1-0.6B Rerank on \textsc{BR-TaxQA} (Figure~\ref{fig:significance_matrices}) are consistent with the semantic matching demands of its question--answer formulation. Taken together, the examples and inferential results reinforce our main empirical conclusion: the benchmark is useful precisely because it captures multiple retrieval regimes, including settings where lexical specificity dominates and others where semantic generalization is more important.

\subsection{Benchmark Limitations}

Although \textsc{JU\'A} is designed to improve comparability in Brazilian legal IR, it also has important limitations. First, the benchmark is intentionally heterogeneous: it combines jurisprudence retrieval, normative retrieval, legislative retrieval, and question-answer retrieval under a common reporting scheme. This breadth is valuable for robustness analysis, but it also means that the overall score aggregates tasks with different query distributions, document structures, and notions of relevance. As a result, strong average performance should not be interpreted as evidence that a model behaves uniformly well across all legal retrieval regimes.

Second, the benchmark combines datasets with varying levels of naturalness and annotation procedures. Some settings rely on direct query-document correspondences, while others use graded relevance. This diversity is useful for stress-testing retrieval methods. Still, it weakens direct comparability across datasets and makes the benchmark better suited to comparative evaluation than to any single absolute notion of legal retrieval quality.

Third, some benchmark design choices deliberately increase task difficulty, such as removing the \texttt{ENUNCIADO} field in \textsc{JurisTCU} and constructing \textsc{JU\'A-Juris} with difficulty-aware sampling. These decisions improve discrimination among retrieval systems, but they also mean that the benchmark should be interpreted as a curated evaluation framework rather than as a direct proxy for all real-world legal search behavior. We view this as a reasonable trade-off, but it is important to state it explicitly.

\section{Conclusions} \label{sec:conclusions}

We introduced \textsc{JU\'A}, a public benchmark for retrieval over Brazilian legal text collections, designed to support reproducible comparison across heterogeneous legal retrieval settings. More broadly, we frame \textsc{JU\'A} not only as a benchmark, but as a continuous evaluation infrastructure for a legal IR landscape that is structurally diverse: jurisprudence retrieval, normative search, legislative retrieval, and question-driven legal retrieval do not reward the same modeling assumptions, and should not be treated as a single homogeneous task.

The experimental results confirm this point. Beyond ranking systems, \textsc{JU\'A} exposes meaningful trade-offs among lexical, dense, and reranking approaches across distinct legal retrieval regimes. BM25 remains highly competitive on collections where lexical overlap, institutional phrasing, and domain-specific terminology remain strong signals. Dense retrieval has statistically supported advantages on \textsc{JU\'A-Juris} for Qwen3-Embedding and Harrier OSS v1-27B and on \textsc{BR-TaxQA} for OpenAI 3-small and Harrier OSS v1-27B. Both Harrier OSS v1 rerankers also have statistically supported gains on these two subsets, although Harrier OSS v1-0.6B reranking has lower NDCG@10 on \textsc{JurisTCU}. Remaining reranking values with incomplete preserved runs are reported descriptively. These differences cannot be reduced to a single notion of retrieval quality: methods that improve position-aware ranking need not improve early-hit behavior or top-10 coverage to the same extent.

We view this heterogeneity as a central strength of the benchmark. It makes \textsc{JU\'A} useful for leaderboard comparison, as well as for studying when dense retrieval helps, when lexical retrieval remains difficult to beat, and how retrieval methods transfer across Brazilian legal domains. In this sense, the benchmark contributes both an evaluation infrastructure and an empirical lens on legal retrieval behavior in Portuguese.

Future work should focus on strengthening the benchmark as a resource rather than only extending the leaderboard. In particular, promising next steps include expanding query diversity, incorporating additional legal collections and evaluation settings that further test transfer across institutions and retrieval regimes, and establishing leakage-free protocols for future domain-adaptation studies. We hope that making \textsc{JU\'A} and its leaderboard public will support more transparent, cumulative, and analytically grounded progress in Brazilian legal IR.

\section*{Supplementary ranked runs and reproducibility}

The supplementary material submitted with this manuscript includes complete BM25 ranked runs for all five benchmark datasets, with top-1000 candidates where applicable and full-corpus rankings for smaller collections. These files enable third-party reranking experiments without reindexing the collections. The supplement also includes the machine-readable output of the paired statistical analysis, including non-significant comparisons, and documentation of the per-query runs eligible for inference. The remaining results are reproducible from the publicly identified model checkpoints, the benchmark data, and the evaluation protocol described in this paper.

\section*{Acknowledgment}
The authors used AI-assisted tools for language revision during manuscript preparation. All technical content, interpretation, and final editorial decisions were reviewed by the authors, who take full responsibility for the article.






\bibliography{sn-bibliography}

\end{document}

%% file: tables/dataset_summary.tex
\begin{sidewaystable*}[p]
\centering
\footnotesize
\begin{tabular}{p{1.4cm}p{3.0cm}p{3.0cm}p{3.0cm}p{3.0cm}p{3.0cm}}
\toprule
Dataset & Retrieval regime & Document type & Query style & Relevance & Role in \textsc{JU\'A} \\
\midrule
\textsc{JU\'A-Juris} & Jurisprudence retrieval & TCU curated jurisprudence excerpts & \textit{Enunciado}-based queries & Binary, exact positive pair & New dataset; train/test benchmark split \\
\textsc{JurisTCU} & Jurisprudence retrieval & TCU jurisprudence excerpts & Real keyword queries + synthetic queries & Expert-verified relevance judgments & Adapted external benchmark \\
\textsc{NormasTCU} & Regulatory / normative retrieval & TCU normative acts & Short keyword queries + richer queries & Three-level graded relevance & Adapted external benchmark \\
\textsc{Ulysses-RFCorpus} & Legislative retrieval & Legislative documents with operational feedback & Real user-oriented legislative queries & Real relevance feedback & Adapted external benchmark \\
\textsc{BR-TaxQA} & Question-driven legal retrieval & Tax answers and linked reference material & FAQ-style tax questions & Graded relevance & Adapted external benchmark \\
\bottomrule
\end{tabular}
\caption{Summary of the datasets considered in this paper's benchmark configuration. \textsc{JU\'A} combines two jurisprudence retrieval settings with broader legislative, regulatory, and question-driven legal retrieval scenarios under a common evaluation protocol.}
\label{tab:jua_dataset_summary}
\end{sidewaystable*}

%% file: tables/leaderboard.tex
\begin{sidewaystable*}[p]
\centering
\footnotesize
\begin{tabular}{clcccccccccccc}
\toprule
\multirow{2}{*}{\#} & \multirow{2}{*}{Model} & \multicolumn{2}{c}{\textsc{JU\'A-Juris}} & \multicolumn{2}{c}{\textsc{JurisTCU}} & \multicolumn{2}{c}{\textsc{NormasTCU}} & \multicolumn{2}{c}{\textsc{Ulysses-RFCorpus}} & \multicolumn{2}{c}{\textsc{BR-TaxQA}} & \multicolumn{2}{c}{Overall} \\
\cmidrule(lr){3-4}\cmidrule(lr){5-6}\cmidrule(lr){7-8}\cmidrule(lr){9-10}\cmidrule(lr){11-12}\cmidrule(lr){13-14}
 &  & N & M & N & M & N & M & N & M & N & M & N & M \\
\midrule
\multicolumn{14}{c}{\textit{Lexical Baseline}} \\
\midrule
1 & BM25 (baseline) & 0.103 & 0.073 & 0.362 & 0.669 & 0.356 & 0.464 & 0.531 & 0.696 & 0.571 & 0.589 & 0.385 & 0.498 \\
\midrule
\multicolumn{14}{c}{\textit{Dense Baselines (General-Purpose)}} \\
\midrule
2 & Qwen3-Embedding-8B & 0.217 & 0.170 & 0.332 & \textbf{0.611} & 0.280 & 0.428 & 0.435 & \textbf{0.703} & 0.799 & 0.828 & 0.413 & \textbf{0.548} \\
3 & Qwen3-Embedding-4B & 0.201 & 0.154 & 0.310 & 0.585 & 0.299 & 0.500 & \textbf{0.526} & 0.677 & 0.776 & 0.803 & \textbf{0.422} & 0.544 \\
4 & Harrier OSS v1-27B & \textbf{0.220} & \textbf{0.171} & \textbf{0.332} & 0.605 & 0.235 & 0.329 & 0.487 & 0.626 & \textbf{0.814} & \textbf{0.846} & \textbf{0.418} & 0.515 \\
5 & Harrier OSS v1-0.6B & 0.139 & 0.107 & 0.281 & 0.577 & \textbf{0.318} & \textbf{0.516} & 0.498 & 0.664 & 0.669 & 0.677 & 0.381 & 0.508 \\
6 & Qwen3-Embedding-0.6B & 0.148 & 0.111 & 0.243 & 0.508 & 0.275 & 0.453 & 0.475 & 0.631 & 0.704 & 0.719 & 0.369 & 0.484 \\
7 & OpenAI 3-small & 0.103 & 0.077 & 0.209 & 0.446 & 0.263 & 0.427 & 0.411 & 0.583 & 0.741 & 0.765 & 0.345 & 0.460 \\
8 & Harrier OSS v1-270M & 0.122 & 0.092 & 0.231 & 0.479 & 0.271 & 0.430 & 0.434 & 0.583 & 0.655 & 0.666 & 0.343 & 0.450 \\
\midrule
\multicolumn{14}{c}{\textit{Rerank Models}} \\
\midrule
9 & Qwen3-Embedding-8B Rerank & 0.221 & 0.172 & \textbf{0.341} & 0.635 & 0.285 & 0.437 & 0.457 & \textbf{0.716} & 0.801 & 0.832 & 0.421 & \textbf{0.558} \\
10 & Qwen3-Embedding-4B Rerank & 0.201 & 0.154 & 0.319 & 0.602 & 0.297 & 0.497 & \textbf{0.537} & 0.684 & 0.774 & 0.801 & \textbf{0.426} & 0.548 \\
11 & Harrier OSS v1-27B Rerank & \textbf{0.222} & \textbf{0.174} & \textbf{0.341} & \textbf{0.636} & 0.242 & 0.333 & 0.510 & 0.650 & \textbf{0.814} & \textbf{0.846} & \textbf{0.426} & 0.528 \\
12 & Harrier OSS v1-0.6B Rerank & 0.144 & 0.110 & 0.292 & 0.592 & \textbf{0.322} & \textbf{0.517} & \textbf{0.523} & 0.680 & 0.671 & 0.679 & 0.390 & 0.516 \\
13 & OpenAI 3-small Rerank & 0.111 & 0.083 & 0.231 & 0.483 & 0.268 & 0.435 & 0.438 & 0.601 & 0.741 & 0.765 & 0.358 & 0.473 \\
\bottomrule
\end{tabular}
\caption{Leaderboard results for the datasets included in this paper's \textsc{JU\'A} benchmark configuration (excluding random baseline). Rows are ordered by Overall MRR within each model segment. Metric abbreviations: N = NDCG@10 and M = MRR@10. Boldface marks the best value in each column within multi-model segments. Overall scores are the mean across available datasets for each model. The leftmost column provides row identifiers used for direct references in the text.}
\label{tab:leaderboard_models}
\end{sidewaystable*}

%% file: tables/reranker_comparison.tex
\begin{table*}[t]
\centering
\footnotesize
\setlength{\tabcolsep}{3.5pt}
\begin{tabular}{lrrrrrr}
\toprule
& \multicolumn{2}{c}{Qwen 4B} & \multicolumn{2}{c}{Harrier 27B} & \multicolumn{2}{c}{Harrier 0.6B} \\
\cmidrule(lr){2-3}\cmidrule(lr){4-5}\cmidrule(lr){6-7}
Dataset & $\Delta$NDCG@10 & $\Delta$MRR@10 & $\Delta$NDCG@10 & $\Delta$MRR@10 & $\Delta$NDCG@10 & $\Delta$MRR@10 \\
\midrule
\textsc{JU\'A-Juris} & $-0.000$ & $-0.000$ & $+0.002$ & $+0.002$ & $+0.006$* & $+0.004$ \\
\textsc{JurisTCU} & $+0.009$ & $+0.017$ & $+0.009$ & $+0.031$ & $+0.012$* & $+0.015$* \\
\textsc{NormasTCU} & $-0.002$ & $-0.003$ & $+0.007$ & $+0.004$ & $+0.002$ & $+0.001$ \\
\textsc{Ulysses} & $+0.011$* & $+0.007$ & $+0.023$* & $+0.024$* & $+0.023$* & $+0.014$* \\
\textsc{BR-TaxQA} & $-0.001$ & $-0.002$ & $+0.001$ & $-0.000$ & $+0.001$ & $+0.001$ \\
\bottomrule
\end{tabular}
\caption{Effect of reranking within each model family. Entries are mean per-query differences between the reranking and first-stage configurations ($\Delta = \mathrm{rerank} - \mathrm{first\mbox{-}stage}$); positive values favor reranking. An asterisk denotes significance at $\alpha=0.05$ after Holm correction across the 30 available metric-level comparisons.}
\label{tab:reranker_comparison}
\end{table*}